\documentclass[12pt,a4paper]{article}

\usepackage[dvips]{graphicx}
\usepackage{amssymb, amsmath}
\usepackage{makeidx}  
\usepackage{multicol} 
\usepackage[bottom]{footmisc}
\usepackage{color}

\usepackage[pagewise]{lineno}
\newtheorem{theorem}{Theorem}%[section]
\newtheorem{lemma}{Lemma}%[section]
\newtheorem{definition}{Definition}
\newtheorem{remark}{Remark}%[section]
\makeindex   

\newcommand{\R}{\mathbb{R}} 
\newcommand{\E}{\mathbb{E}}

\begin{document}

\title{
Pricing American Call Options by the Black-Scholes Equation with a Nonlinear Volatility Function
}

\author{Maria do Ros\'ario Grossinho, Yaser Faghan Kord \thanks{Instituto Superior de Economia e Gest\~ao and CEMAPRE, Universidade de Lisboa, Portugal.\tt{mrg@iseg.ulisboa.pt, yaser.kord@yahoo.com}} and Daniel \v{S}ev\v{c}ovi\v{c}\thanks{Dept.\ Applied Mathematics \& Statistics, Comenius University, 842 48  Bratislava, Slovakia.\tt{sevcovic@fmph.uniba.sk}}}

\maketitle

\begin{abstract}
In this paper we investigate a nonlinear generalization of the Black-Scholes equation for pricing American style call options in which  the volatility term may depend on the underlying asset price and the Gamma of the option. We propose a numerical method for pricing American style call options  by means of transformation of the free boundary problem for a nonlinear Black-Scholes equation into the so-called Gamma variational inequality with the new variable depending on the Gamma of the option. We apply a modified projective successive over relaxation method in order to construct an effective numerical scheme for discretization of the Gamma variational inequality. Finally, we present several computational examples for the nonlinear Black-Scholes equation for pricing American style call option under presence of variable transaction costs. 

\end{abstract}

\noindent\textit{Keywords and phrases:}{
American option pricing, nonlinear Black-Scholes equation, variable transaction costs, PSOR method}

\noindent\textit{Mathematics Subject Classification:}{
35K15, 35K55, 90A09, 91B28.}

%%%%%%%%%%%%%%%%%%%%%%%%%%%%%%%%%%%%%%%%%%%%%%%%%%%%%%%%%%%%

\section{Introduction}

In a stylized financial market, the price of a European style option can be computed from a solution to the well-known Black--Scholes linear parabolic equation derived by Black and Scholes in \cite{BS}. Recall that a European call option gives its owner the right but not obligation to purchase an underlying asset at the expiration price $E$ at the expiration time $T$.
In this paper, we consider American style options which can be exercised at anytime $t$ in the time interval $[0,T]$.

The classical linear Black Scholes model was derived under several restrictive assumptions, namely no transaction costs, frictionless, liquid and complete market, etc. However, we need more realistic models for the market data analysis in order to overcome the drawbacks due to these restrictions of the classical Black-Scholes theory. One of the first nonlinear models taking into account transaction costs is the jumping volatility model by Avellaneda,  L\'evy and Paras \cite{AP}.
The nonlinearity of the original Black-Scholes model can also arise from the feedback and illiquid market effects due to large traders choosing given stock-trading
strategies (Sch\"onbucher and Wilmott \cite{SW}, Frey and Patie \cite{FP}, Frey and Stremme \cite{FS}), imperfect replication and investor’s preferences (Barles and Soner \cite{BaSo}), risk from unprotected portfolio (Kratka \cite{Kr}, Janda\v{c}ka and \v{S}ev\v{c}ovi\v{c} \cite{JS}). In this paper we are concerned with a new nonlinear model derived recently by \v{S}ev\v{c}ovi\v{c} and \v Zit\v{n}ansk\'a \cite{SeZ} for pricing call or put options in the presence of variable transaction costs. The model generalizes the well-known Leland model with constant transaction costs function (cf. \cite{Le}, \cite{HWW}) and the Amster et al. model \cite{AAMR} with linearly decreasing transaction costs function. It leads to the following generalized Black-Scholes equation with the nonlinear volatility function $\hat\sigma$ depending on the product $H=S \partial^2_S V$ of the underlying asset price $S$ and the second derivative (Gamma) of the option price $V$:
\begin{equation}
\partial_t V + \frac{1}{2} \hat\sigma(S\partial _S^2V)^2 S^2 \partial^2_S V + (r-q) S \partial_S V -r V =0, \quad  V(T,S) = (S-E)^+,
\label{nonlinear-BS}
\end{equation} 
where $r,q\ge 0$ are the interest rate and the dividend yield, respectively. The price $V(t,S)$ of a such a call option, in the presence of variable transaction costs, is given by a solution to the nonlinear parabolic equation (\ref{nonlinear-BS}) depending on the underlying stock price $S>0$ at the time $t\in[0,T]$, where $T>0$ is the time of maturity and $E>0$ is the exercise price. 

For European style call options various numerical methods for solving the fully nonlinear parabolic equation (\ref{nonlinear-BS}) were proposed and analyzed by \v{D}uri\v{s} et al. in \cite{DTLS}. In \cite{Se2} and \cite{SeZ} \v{S}ev\v{c}ovi\v{c}, Janda\v{c}ka  and \v Zit\v{n}ansk\'a  investigated a new transformation technique (referred to as the Gamma transformation). They showed that the fully nonlinear parabolic equation (\ref{nonlinear-BS}) can be transformed to a quasilinear parabolic equation 
\begin{equation}\label{Y2intro}
\partial_\tau H-\partial^2_u\beta(H)-\partial_u\beta(H)-(r-q)\partial_u H+qH=0, \quad \hbox{where}\ \beta(H)=\hat\sigma(H)^2 H/2, 
\end{equation} 
of a porous-media type for the transformed quantity $H(\tau,u) = S\partial_S^2 V(t,S)$ where $\tau=T-t$, $u=\ln(S/E)$.

The advantage of solving the quasilinear parabolic equation in the divergent form (\ref{Y2intro}) compared to the fully nonlinear equation (\ref{nonlinear-BS}) is twofold. Firstly, from the analytical point of view, the theory of existence, uniqueness of solutions to quasilinear parabolic equation of the form (\ref{Y2intro}) is well developed and understood. Using the general theory of quasilinear parabolic equations due to  Ladyzhenskaya et. al \cite{LSU}, the existence of H\"older smooth solutions to (\ref{Y2intro}) has been shown in \cite{SeZ} by \v{S}ev\v{c}ovi\v{c} and \v Zit\v{n}ansk\'a. Secondly, the quasilinear parabolic equations in the divergent form can be numerically approximated by means of the finite volume method (cf. LeVeque \cite{LeV}). Furthermore, the semi-implicit approximation scheme proposed in Section 4 fits into a class of methods which have been shown to be of the second order of convergence (see e.g. Kilianov\'a and \v{S}ev\v{c}ovi\v{c} \cite{KS}). In a series of papers \cite{Kole1,Kole2,Kole3,Kole4} Koleva and Vulkov investigated the transformed Gamma equation (\ref{Y2intro}) for pricing European style of call and put options. They also derived the second order positivity preserving numerical scheme for solving (\ref{nonlinear-BS}) and (\ref{Y2intro}).

Our goal is to study American style call options which, as known,  leads to a free boundary problem. Their prices can be computed by means of the generalized Black-Scholes equation with the nonlinear volatility function (\ref{nonlinear-BS}). If the volatility function is constant then it is well known that American options can be priced by means of a solution to a linear complementarity problem (cf. Kwok \cite{Kw}). Similarly, for the nonlinear volatility model, one can construct a nonlinear complementarity problem involving the variational inequality for the left-hand side of (\ref{nonlinear-BS}) and the inequality $V(t,S)\ge (S-E)^+$. However, due to the fully nonlinear nature of the differential operator in (\ref{nonlinear-BS}), the direct computation of the nonlinear complementarity becomes harder and unstable. Therefore, we propose an alternative approach and reformulate the nonlinear complementarity problem in terms of the new transformed variable $H$ for which the differential operator has the form of a quasilinear parabolic operator appearing in the left-hand side of (\ref{Y2intro}).

In order to apply the Gamma transformation for American style options we derive the nonlinear complementarity problem for the transformed variable $H$ and we solve the variational problem by means of the modified projected successive over relaxation method  (cf. Kwok \cite{Kw}). Using this method we compute American style call option prices for the Black-Scholes nonlinear model for pricing call options in the presence of variable transaction costs. 
 
The paper is organized as follows. In section $2$, we present a nonlinear option pricing model under variable transaction costs. Section $3$ is devoted to the transformation of the free boundary problem into the so-called Gamma variational inequality. In section $4$, we present a finite volume discretization of the complementarity problem and its solution obtained by means of the projected super over relaxation (PSOR) method. Finally, in section $5$, we present results of various numerical experiments for pricing American style of call options, the early exercise boundary position and comparison with models with constant volatility terms.

\section{Nonlinear Black-Scholes equation for pricing options in the presence of variable transaction costs}

In the original Black-Scholes theory, continuous hedging of the portfolio including underlying stocks and options is allowed. In the presence of transaction costs for purchasing and selling the underlying stock, this continuous feature may lead to an infinite number of transaction costs yielding unbounded total transaction costs.   

One of the basic nonlinear models including transaction costs is the Leland model \cite{Le} for  option pricing in which the possibility of rearranging portfolio at discrete time can be relaxed. Recall that, in the derivation of the Leland model \cite{HWW,H,Le}, it is assumed that the investor follows the delta hedging strategy in which the number $\delta$ of bought/sold underlying assets depends on the delta of the option, i.e. $\delta=\partial_S V$. Then, applying self-financing portfolio arguments, one can derive the extended version of the Black Scholes equation    
\begin{equation}
\label{Y3.10}
 \partial_tV+(r-q)S\partial_S V+\frac{1}{2}\sigma^2 S^2\partial_S^2 V- r V = r_{TC}S. 
\end{equation}
Here the  transaction cost measure $r_{TC}$ is given by
\begin{equation}
r_{TC}=\frac{\mathbb{E}[\Delta TC]}{S\Delta t},
\end{equation}  
where $\Delta TC$ is the change in transaction costs during the time interval of the length $\Delta t>0$. If $C\ge 0$ represents a percentage of the cost of the sale and purchase of a share relative to the price $S$ then $\Delta TC = \frac12 C S |\Delta\delta|$ where  $\Delta\delta$ is the number of bought ($\Delta\delta>0$) or sold ($\Delta\delta<0$) underlying assets during the time interval $\Delta t$.  The parameter $C>0$ measuring transaction costs per unit of the underlying asset can be either constant or it may depend on the number of transacted underlying assets, i.e. $C=C(\left|\Delta\delta\right|)$.

Furthermore, assuming the underlying asset follows the geometric Brownian motion $dS = \mu S dt + \sigma S d W$ it can be shown that $\Delta \delta = \Delta \partial_SV \approx \sigma S \partial_S^2V \Phi \sqrt{\Delta t}$ where  $\Phi\sim N(0,1)$ is normally distributed random variable. Hence
\begin{equation}
\label{Y3.7}
r_{TC}=\frac{1}{2}\frac{\mathbb{E}[C(\alpha\left|\Phi\right|)\alpha\left|\Phi\right|]}{\Delta t},
\end{equation} 
where $\alpha:= \sigma S |\partial_S^2V|\sqrt{\Delta t}$ (cf.  \cite{SeA}, \cite{JS}). In order to rewrite equation (\ref{Y3.10}) we recall the notion of the  mean value modification of the transaction cost function introduced in \cite{SeZ}).

\begin{definition}\cite[Definition 1]{SeZ}
\label{def:tilde_C}
Let $C=C(\xi)$,  $C:\R^+_0 \to \R$, be a transaction costs function. The integral transformation $\tilde{C}:\R^+_0 \to \R$ of the function $C$ defined as follows:
\begin{equation}\label{r:tildeC}
\tilde C(\xi) = \sqrt{\frac{\pi}{2}}\E [C(\xi|\Phi|)|\Phi|] 
= \int_0^\infty C(\xi x) x\, e^{-x^2/2}  dx,
\end{equation}
is called the mean value modification of the transaction costs function. Here $\Phi$ is the random variable with a standardized normal distribution, i.e., $\Phi\sim N(0,1)$.
\end{definition}

If we assume that $C:\R^+_0\rightarrow\R$ is a measurable and bounded transaction costs function then the price of the option based on the variable transaction costs is given by the solution of the following nonlinear Black-Scholes PDE (cf. \cite[Proposition 2.1]{SeZ}):
\begin{equation}
\label{Y.ttt}
 \partial_tV+(r-q)S\partial_S V+\frac{1}{2}\hat{\sigma}(S\partial_S^2 V)^2S^2\partial_S^2 V-rV=0,
 \end{equation}   
where the nonlinear diffusion coefficient $\hat{\sigma}^2$ is given by (denoting $\hat{\sigma}^2(.):=\hat{\sigma}(.)^2)$:
\begin{equation}
\hat{\sigma}(S\partial _S^2 V)^2=\sigma^2\left(1-\sqrt{\frac{2}{\pi}}\tilde{C}(\sigma S\left|\partial _S^2V\right|\sqrt{\Delta t})\frac{\mathrm{sgn}({S\partial _S^2V})}{\sigma\sqrt{\Delta t}}\right).
\end{equation} 

\begin{figure}
\begin{center}
\includegraphics[width=0.45\textwidth]{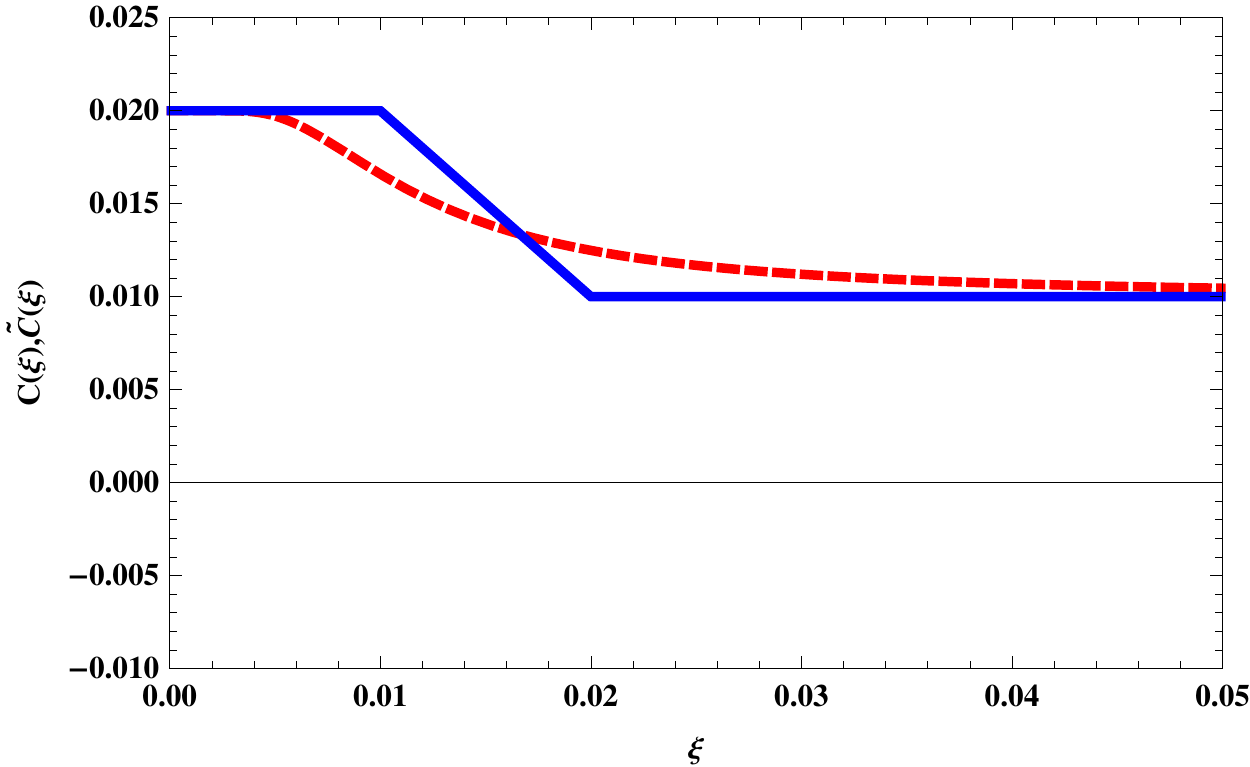}
\end{center}
\caption{%
A piecewise linear transaction costs function with parameters $C_0 = 0.02, \kappa = 1, \xi_- = 0.01, \xi_+ = 0.02$ and its mean value modification $\tilde C(\xi)$ (dashed line).}
\label{fig:Cfun}
\end{figure}

A realistic example of a transaction costs function was proposed and analyzed by \v{S}ev\v{c}ovi\v{c} and \v Zit\v{n}ansk\'a in \cite{SeZ}. A piecewise linear decreasing transaction costs function $C$ is given by 
\begin{equation}
C(\xi)=\left\{
\begin{tabular}{lcl}
\label{Y3.18}
$C_0$, & if & $0\leq\xi<\xi_{-}$,\\
$C_0-\kappa(\xi-\xi_{-})$, & if & $\xi_{-}\leq \xi \leq \xi_{+}$,\\
	 $\underline{C_0}\equiv C_0-\kappa(\xi_{+}-\xi_{-})$, & if & $\xi\geq\xi_{+}$,
\end{tabular}
\right.
\end{equation} 
where $0<\xi_{-}\leq\xi_{+}, \kappa>0, C_0>0$ are model parameters. Such a transaction costs function corresponds to a stylized market in which the investor pays the amount $C_0$ for a small volume of traded assets whereas, if the traded volume of stocks is higher, the investor pays a smaller amount $\underline{C_0}$. The modified mean value transaction costs function can be analytically expressed by the following formula:
\begin{equation}
\label{Y3.17}
\tilde{C}(\xi)=C_0 - \kappa\xi\int_\frac{\xi_{-}}{\xi}^\frac{\xi_{+}}{\xi}e^{-u^2/2} du, \quad \text{for} \quad \xi\geq0
\end{equation}
(cf. \cite[Eq. (24)]{SeZ}). According to \cite[Proposition 2.2]{SeZ} there exist lower/upper bounds and limiting behavior of the mean value modification of the piecewise linear transaction costs function $\tilde{C}(\xi)$, i.e. 
\begin{equation}
\underline{C_0} \leq\tilde{C}(\xi)\leq C_0, \quad\hbox{and}\ \
\lim_{\xi\rightarrow\infty} \tilde{C}(\xi)=\lim_{\xi\rightarrow\infty} C(\xi)=\underline{C_0}. 
\end{equation}
A graph of a piecewise linear transaction costs function $C$ and its mean value modification is depicted in Fig.~\ref{fig:Cfun}.

In the case when the transaction cost function $C\equiv C_0>0$ is constant (i.e. $\xi_\pm=0$) we obtain the well-known Leland model (cf. \cite{HWW, H, Le}) in which the diffusion term has the form:
\[
\hat{\sigma}(S\partial_S^2 V)^2=\sigma^2(1- \mathrm{Le}\, {\mathrm{sgn}({\partial _S^2V})})= \sigma^2(1- \mathrm{Le}\, {\mathrm{sgn}({S\partial _S^2V})}),\quad \mathrm{Le}=\sqrt{\frac{2}{\pi}}\frac{C_0}{\sigma \sqrt{\Delta t}},
\]
where $\mathrm{Le}$ is the Leland number. 

In \cite{AAMR} Amster et al. investigated a linear non-increasing transaction costs function of the form:
\[
C(\xi)=C_0-\kappa \xi, \quad \text{where} \quad \xi\geq0,
\]
i.e. $\xi_-=0, \xi_+=\infty$. The mean value modification function has the form:
\[
\tilde{C}(\xi)=C_0 - \sqrt{\pi/2}\kappa  \xi, \quad \text{where} \quad \xi\geq0.
\]
Clearly, such a transaction costs function can attain negative values which can be considered as a drawback of this model.

%%%%%%%%%%%%%%%%%%%%%%%%%%%%%%%%%%%%%%%%%%%%%%%%%%%%%%

\section{Transformation of the free boundary problem to the Gamma variational inequality}

In the context of European style options the transformation method to the Gamma equation was proposed and analyzed by Janda\v{c}ka and \v{S}ev\v{c}ovi\v{c} \cite{JS}. If we consider the generalized nonlinear Black-Scholes equation (\ref{nonlinear-BS}) for the European style of an option then, making the change of variables $u=\ln(\frac{S}{E})$ and $\tau=T-t$ and  computing the second derivative of equation (\ref{nonlinear-BS}) with respect to $u$, we derive the so-called Gamma equation (\ref{Y2intro}), i.e.
\begin{equation}\label{Y2}
\partial_\tau H-\partial_u\beta(H)-\partial^2_u\beta(H)-(r-q)\partial_u H+qH=0, \quad \hbox{where} \ \ \beta(H)=\frac{1}{2}\hat\sigma(H)^2 H .
\end{equation}
More details of derivation of the Gamma equation, existence and uniqueness of classical H\"older smooth solutions can be found in the paper  \cite{SeZ} by  \v{S}ev\v{c}ovi\v{c} and \v Zit\v{n}ansk\'a. 

\begin{lemma}\cite[Proposition 3.1, Remark 3.1]{SeZ}
Let us consider a call option with the pay-off diagram $V(T,S)=(S-E)^+$. Then the function $H(\tau,u)=S\partial^2_SV(t,S)$ where $u=\ln(\frac{S}{E})$ and $\tau=T-t$ is a solution to (\ref{Y2}) subject to the Dirac initial condition $H(0,x)=\delta(x)$ if and only if 
\begin{equation*}\label{Y2V}
V(t,S)=\int_{-\infty}^{+\infty} (S-Ee^{u})^+ H(\tau,u) du
\end{equation*} 
is a solution to (\ref{nonlinear-BS}). 
\end{lemma}

\subsection{American style options}

In this section we investigate the transformation method of a free boundary problem arising in pricing American style of options by means of a solution to the so-called Gamma variational inequality. 

The principal advantage of American style of an option contract is the flexibility it offers to its holder as it can be exercised anytime before the expiration date $T$. Majority derivative contracts traded in financial markets are of the American style. In mathematical modeling of American options, unlike European style options, there is the possibility of early exercising the contract at some time $t^{*}\in [0,T)$ prior to the maturity time $T$. 

It is well-known that  pricing an American call option on an underlying stock paying continuous dividend yield $q>0$ leads to a free boundary problem. In addition to a function $V(t,S)$, we need to find the early exercise boundary function $S_{f}(t), \ t\in [0,T]$.  The function $S_{f}(t)$ has the following properties:
\begin{itemize}
\item If $S_{f}(t)>S$ for $t\in[0,T]$ then $V(t,S)>(S-E)^+$.
\item If $S_{f}(t)\leq S$ for $t\in[0,T]$ then $V(t,S)=(S-E)^+$.   
\end{itemize}  

In the last decades many authors analyzed the free boundary position function $S_f$. In \cite{SSC} Stamicar, \v{S}ev\v{c}ovi\v{c} and Chadam derived accurate approximation to the early exercise position for times $t$ close to expiry $T$ for the Black-Scholes model with constant volatility (see also \cite{KK}, \cite{LS}, \cite{SPZHU}). The method has been generalized for the nonlinear Black-Scholes model by \v{S}ev\v{c}ovi\v{c} in \cite{Se2}.

Following Kwok \cite{Kw} (see also \cite{SeA}) the free boundary problem for pricing an American call option consists in finding a function $V(t,S)$ and the early exercise boundary function $S_f$ such that $V$ solves the Black-Scholes PDE (\ref{nonlinear-BS}) on a time depending domain: $\{(t,S), \  0<S <S_{f}(t)\}$ and $V(t,S_{f}(t)) =S_{f}(t)-E$, and $\partial_S  V (t,S_{f}(t)) = 1$.

Alternatively, a $C^1$ smooth function $V$ is a solution to the free boundary problem for (\ref{nonlinear-BS}) if and only if it is a solution to the nonlinear variational inequality  
\begin{eqnarray}
&&\partial_t V+(r-q)S\partial_S V+ S \beta(S \partial^2_S V)-rV\leq0,
\qquad V(t,S)\geq g(S),
\nonumber
\\
&& \left(\partial_t V+(r-q)S\partial_S V+ S \beta(S \partial^2_S V)-rV\right) \times \left(V  - g \right) = 0,
\label{NLCP}
\end{eqnarray}
for any $S>0$ and $t\in[0,T]$ where $g(S)\equiv(S-E)^{+}$.

%%%%%%%%%%%%%%%%%%%%%%%%%%%%%%%%%%%%%%%%%%%%%%%%%%%%%%%%%%%%

\subsection{Gamma transformation of the variational inequality}

In this section we present a transformation technique how to transform the nonlinear complementarity problem (\ref{NLCP}) for the function $V(t,S)$ into the so-called Gamma variational inequality involving the transformed function $H(\tau,x)$. We need two auxiliary lemmas. 

\begin{lemma}\label{Y00}
Let $V(t,S)$ be a given function. Let $u=\ln(\frac{S}{E})$, $\tau=T-t$ and define the function $Y(\tau,u):=\partial_t V+(r-q)S\partial_S V+ S \beta(S\partial_S^2 V)-rV$. Then 
\begin{equation*}
-\partial_\tau H+\partial_u\beta(H)+\partial^2_u\beta(H)+(r-q)\partial_u H-qH=\frac{1}{E}e^{-u}[\partial_u^2 Y-\partial_u Y],
\end{equation*} 
where $H(\tau,u)=S \partial^2_S V(t,S)$.
\end{lemma}

\noindent {P r o o f.} 
By differentiating the function $Y$ with respect to the  $u$ variable and using the fact that $\partial_u=S\partial_S$, we obtain 
\begin{eqnarray}
\partial_u Y &=& \partial_t(S\partial_S V) + S(\beta+\partial_u \beta)+ (r-q)S H-qS\partial_S V,
\nonumber
\\
\partial_u^2 Y &=& \partial_t(S\partial_S V+S^2\partial_S^2 V)+ (r-q)S(H+\partial_u H)
\label{Y3003}
\\
&& +S(\beta+\partial_u \beta)+S(\partial_x^2\beta+\partial_u \beta)-qS\partial_S V-qH,
\nonumber
\end{eqnarray}
where $S=Ee^{u}$. Then $\partial_u^2 Y-\partial_u Y = Ee^{u}\Psi[H]$, where 
\begin{equation}
\Psi[H]:=-\partial_\tau H+\partial_u\beta(H)+\partial^2_u\beta(H)+(r-q)\partial_u H-q H,
\label{Y77}
\end{equation}
as claimed. \hfill $\diamondsuit$

\medskip

In the particular case  when $Y\equiv0$, the function $V(t,S)$ represents a price of the European style call option  and it is a solution to the nonlinear Black-Scholes equation (\ref{nonlinear-BS}) if and only if the function $H$ is a solution to the so-called Gamma equation (\ref{Y2}) subject to the initial condition $H(x,0)=\delta(x)$, where $\delta$ is the Dirac function  (cf. \cite{Se2}, \cite{SeZ}).

\begin{lemma}\label{Y111}
If the function $Y$ fulfills the asymptotic behavior 
$\lim_{u\to{-\infty}}Y(\tau,u)=0$ and $\lim_{u\to{-\infty}}e^{-u}\partial_u Y(\tau,u)=0$ then
\begin{equation*}
\int_{-\infty}^{+\infty} (S-Ee^{u})^+\Psi[H](\tau,u) du =Y(\tau,u)|_{u=\ln(S/E)} \equiv \partial_t V+(r-q)S\partial_S V+ S \beta(S\partial_S^2 V)-rV.
\end{equation*} 
\end{lemma}

\noindent {P r o o f.} Using Lemma 2 and (\ref{Y3003}) we can express the term $\int_{-\infty}^{+\infty} (S-Ee^{u})^+\Psi[H](\tau,u) du$ as follows: 
\begin{eqnarray*}
&& \int_{-\infty}^{+\infty} (S-Ee^{u})^+ \frac{1}{E}e^{-u}[\partial_u^2 Y-\partial_u Y] du 
=\frac{1}{E}\int_{-\infty}^{\ln(S/E)} (Se^{-u}-E) [\partial_u^2 Y-\partial_u Y] du
\\
&&= \frac{1}{E}\int_{-\infty}^{\ln(S/E)} \left[Se^{-u}\partial_u Y-(Se^{-u}-E)\partial_u Y du\right] +\underbrace{[(Se^{-u}-E)\partial_u Y]_{-\infty}^{\ln(S/E)}}_{0}
\\
&&=\frac{1}{E}\int_{-\infty}^{+\infty}E\partial_u Y du =Y(\tau,u)|_{u=\ln(S/E)} =\partial_t V+(r-q)S\partial_S V+ S \beta(S\partial_S^2 V)-rV,
\end{eqnarray*}
and proof of lemma follows. \hfill $\diamondsuit$

\begin{theorem}\label{Y0000}
The function $V(t,S)$ is a solution to the nonlinear complementarity problem (\ref{NLCP}) if and only if the transformed function $H$ is a solution the following Gamma variational inequality and complementarity constraint:
\begin{eqnarray}
\label{Y5.20}
&&-\int_{-\infty}^{+\infty}(S-Ee^{u})^+ \Psi[H](\tau,u) du\geq0, \quad 
\int_{-\infty}^{+\infty}(S-Ee^{u})^+H(\tau,u) du \geq g(S), 
\\
\label{Y5.21-LC}
&&\int_{-\infty}^{+\infty}(S-Ee^{u})^+ \Psi[H](\tau,u)) du
\times \left(\int_{-\infty}^{+\infty}(S-Ee^{u})^+H(\tau,u) du - g(S) \right) =0, \nonumber 
\end{eqnarray}
for any $S\geq0$ and $\tau\in[0,T]$.
\end{theorem}

\medskip
\noindent {P r o o f.} It directly follows from Lemma~\ref{Y00} and Lemma~\ref{Y111}. 
\hfill $\diamondsuit$

\medskip

\begin{remark}
For calculating $V(T,S)$ in Theorem~\ref{Y0000} we use the fact that $H(0,u)=\bar{H}(u), u\in \R$, where $\bar{H}(u):=\delta(u)$ is the Dirac delta function such that $\int_{-\infty}^{+\infty}\delta(u)du=1$, and $\int_{-\infty}^{+\infty}\delta(u-u_0)\phi(u) du=\phi(u_0)$ 
for any continuous function $\phi$. 

In what follows, we will approximate the initial Dirac $\delta$-function as follows:
\[
H(x,0) \approx f(d)/(\hat\sigma\sqrt{\tau^*}),
\] 
where $0<\tau^*\ll 1$ is a sufficiently small parameter, $f(d)$ is the PDF function of the normal distribution, that is: $f(d) =e^{-d^2/2}/\sqrt{2\pi}$ and $d=\left(x+ (r-q-\sigma^2/2)\tau^* \right)/\sigma\sqrt{\tau^*}$. This approximation follows from observation that for a  solution of the linear Black--Scholes equation with a constant volatility $\sigma>0$ at the time $T-\tau^*$ close to expiry $T$ the value $H^{lin}(x,\tau^*) = S\partial^2_S V^{lin}(S, T-\tau^*)$ is given by  $H^{lin}(x,\tau^*) = f(d)/(\hat\sigma\sqrt{\tau^*})$. Moreover,  $H^{lin}(.,\tau^*) \to \delta(.)$ as $\tau^*\to0$ in the sense of distributions. 
\end{remark}

%%%%%%%%%%%%%%%%%%%%%%%%%%%%%%%%%%%%%%%%%%%%%%%%%%%%%%%%%%%%

\section{Solving the Gamma variational inequality by the PSOR method}
According to Theorem~\ref{Y0000}, the American call option pricing problem can be rewritten in terms of the function $H(\tau, u)$ in the form of the Gamma variational inequality (\ref{Y5.20}) with the complementarity constraint (\ref{Y5.21-LC}). We follow the paper by \v{S}ev\v{c}ovi\v{c} and \v Zit\v{n}ansk\'a \cite{SeZ} in order to derive an efficient numerical scheme for solving the Gamma variational inequality for a general form of the function $\beta(H)$ including the special case of the variable transaction costs model.  In order to apply the Projected Successive Over Relaxation method (PSOR) (cf. Kwok \cite{Kw}) to the variational inequality (\ref{Y5.20}), we have to discretize the nonlinear operator $\Psi$ defined in (\ref{Y77}). 

The proposed numerical discretization is based on the finite volume method. Assume that the spatial variable $u$ belongs to the bounded  interval $(-L,L)$ for sufficiently large $L>0$. We divide the spatial interval $[-L,L]$ into a uniform mesh of discrete points $u_i=ih$ where $i=-n,\cdots,n$ with a spatial step $h=\frac{L}{n}$. The time interval $[0,T]$ is uniformly divided with a time step $k=\frac{T}{m}$ into discrete points $\tau_j=jk$ for $j=1,\cdots,m$. The finite volume discretization of the operator $\Psi[H]$ leads to a tridiagonal matrix multiplied by the vector $H^j=(H^j_{-n+1},\cdots,H^j_{n-1})\in\R^{2n-1}$. More precisely, the vector $\Psi[H]^j$ at the time level $\tau_j$ is given by  $\Psi[H]^j= -(A^{j}H^j-d^j)$ where the $(2n-1)\times(2n-1)$ matrix $A^j$ has the form 
\begin{equation}\label{Y5.888}
A^j = 
 \begin{pmatrix}
  b_{-n+1}^j & c_{-n+1}^j & 0 & \cdots & 0 \\
  a_{-n+2}^j & b_{-n+2}^j & c_{-n+2}^j &  & \vdots \\
  0  & .  & . & .& 0  \\
  \vdots & \cdots & a_{n-2}^j & b_{n-2}^j & c_{n-2}^j \\
  0 &\cdots & 0 & a_{n-1}^j & b_{n-1}^j
 \end{pmatrix}
\end{equation}
with coefficients:  
\begin{eqnarray*}
a_i^j&=&-\frac{k}{h^2}\beta'(H_{i-1}^{j-1})+\frac{k}{2h}(r-q), \\
c_i^j&=&-\frac{k}{h^2}\beta'(H_i^{j-1})-\frac{k}{2h}(r-q),\\
b_i^j&=&(1+kq)-(a_i^j+c_i^j),\\
d_i^j &=&  H_i^{j-1}+\frac{k}{h}\left(\beta(H_i^{j-1})-\beta(H_{i-1}^{j-1})\right).
\end{eqnarray*}
Finally, using simple numerical integration the variational inequality (\ref{Y5.20})  can be discretized as follows:
\begin{equation} \label{Y5.66}
V(S,T-\tau_j)=h\sum_{i=-n}^{n}(S-Ee^{u_i})^+ H^j_i, \quad j=1,2,\cdots,m. 
\end{equation}         
Then, the full space-time discretized version of the inequalities occurring in (\ref{Y5.20}) is given by  
\begin{eqnarray}
\label{Y5.4}
h\sum_{i=-n}^{n}(S-Ee^{u_i})^+\left[(A^jH^j)_i- d^j_i\right]&\geq&0,\\
\label{Y5.5}
h\sum_{i=-n}^{n}(S-Ee^{u_i})^+H_i^j&\geq& g(S)\equiv (S-E)^+.
\end{eqnarray}
Let us define the auxiliary invertible matrix $P=(P_{li})$ as follows:
\begin{eqnarray} \label{Y5.55} 
P_{li}&=&h\, \max(S_l-Ee^{u_i},0) =h E\, \max(e^{v_l}-e^{u_i},0) 
\end{eqnarray}
where $v_l=(u_{l+1}+u_{l-1})/2$, for $l=-n, \cdots, n$.

Next, our purpose is to solve the problem (\ref{Y5.4})--(\ref{Y5.5}) by means of the PSOR method. Using the matrix $P$  we can rewrite (\ref{Y5.4})--(\ref{Y5.5}) for the American call option as follows
\begin{eqnarray}
&& (P A H)_i\geq (P d)_-, \quad (P H)_i\geq g_i, 
\label{Y5.6}
\\
&&(PAH-Pd)_i\cdot (PH-g)_i=0,\ \ \hbox{for all}\  i,
\nonumber
\end{eqnarray} 
where $A=A^j$, $g_i= (S_i-E)^+$ and $H=H^j$. The complementarity problem (\ref{Y5.6}) can be solved by means of the PSOR algorithm, given by the following iterative scheme:
\begin{enumerate}
 \item for $k=0$ set $v^{j,k}=v^{j-1}$,
 \item until $k\le k_{max}$ repeat:
\begin{eqnarray*}
 w^{j, k+1}_i &=&\frac{1}{\tilde A_{ii}}\left(-\sum_{l<i}\tilde A_{il} v^{j,k+1}_l-\sum_{l>i}\tilde A_{il} v^{j,k}_l + \tilde d^j_i\right),\\
 v^{j,k+1}_i&=&\max\left\{v^{j,k}_i +\omega (w^{j,k+1}_i - v^{j,k}_i),\,  g_i\right\},
\end{eqnarray*}
\item set $v^j= v^{j,k+1}$,
\end{enumerate}
for $i=-n, \cdots,n$ and $j=1, \cdots,m$, where $v^j=P H^j$, $\tilde d^j=P d^j$  and $\tilde A = P A^j P^{-1}$. Here $\omega\in[1,2]$ is a relaxation parameter which can be tuned in order to speed up the  convergence process. Finally, using the value $H^j=P^{-1} v^j$ and equation (\ref{Y5.66}), we can evaluate the option price $V$.

%%%%%%%%%%%%%%%%%%%%%%%%%%%%%%%%%%%%%%%%%%%%%%%%%%%%%%%

\section{Numerical experiments}

\begin{figure}
\begin{center}
\includegraphics[width=0.4\textwidth]{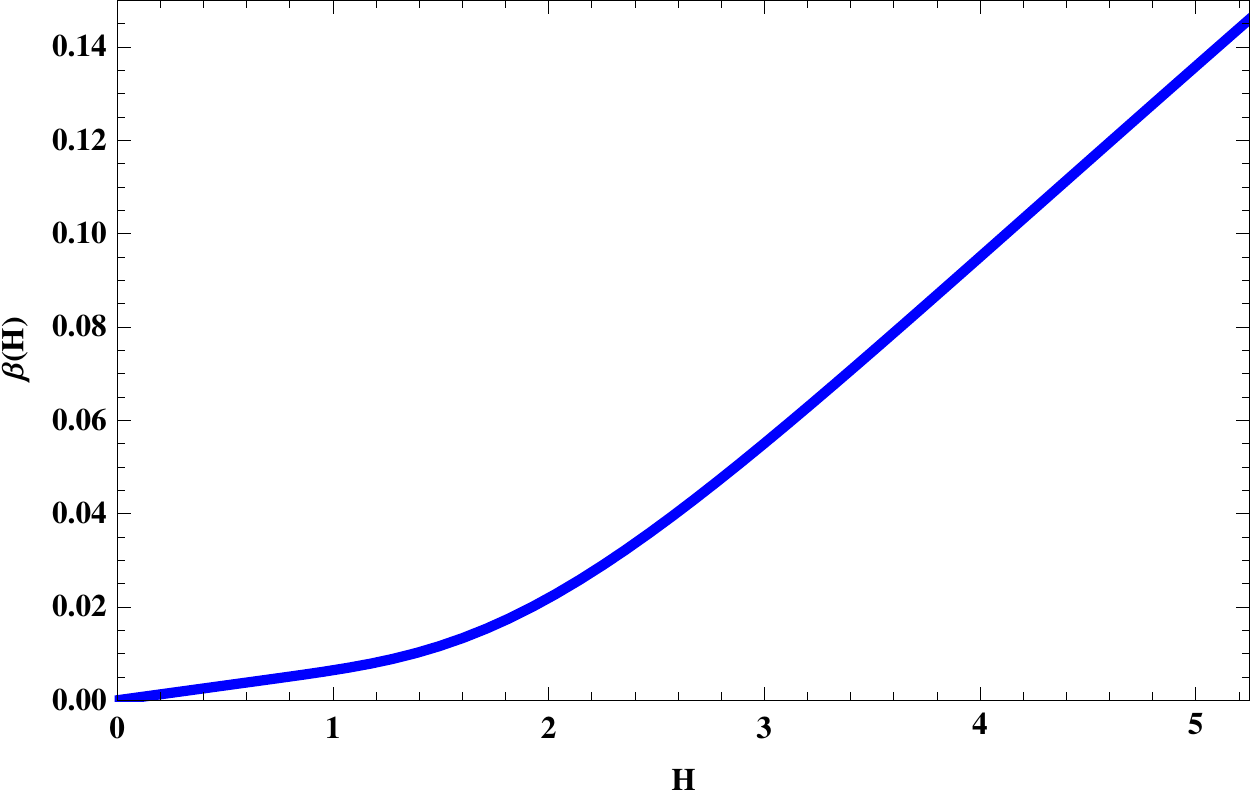}
\end{center}

\caption{A graph of the function $\beta(H)$ related to the piecewise linear decreasing transaction costs function (see \cite{JS}).}
\label{beta}
\end{figure}

In this section, we focus our attention on numerical experiments for computing American style  call option prices based on the nonlinear Black-Scholes equation that includes a piecewise linear decreasing transaction costs function $C$. In  Fig.~\ref{beta}, we show the corresponding function $\beta(H)$ given by 
\begin{equation*}
\beta(H)=\frac{\sigma^2}{2}\left(1- \sqrt{\frac{2}{\pi}} \tilde{C}(\sigma |H|\sqrt{\Delta t})\frac{\mathrm{sgn}(H)}{\sigma\sqrt{\Delta t}}\right)H,
\end{equation*}          
where $\tilde{C}$ is the mean value modification of the transaction costs function $C$.

The parameters $C_0, \kappa, \xi_\pm, \Delta t$ characterizing the nonlinear piecewise linear variable transaction costs function $C$ and other model parameters are presented in Table~\ref{table:1}. Here  $\Delta t$ is the time interval between two consecutive portfolio rearrangements,  $T$ is the maturity time, $\sigma$ is the historical volatility, $q$ is the dividend yield, $E$ is the strike price and $r$ denotes the risk free interest rate. The small parameter $0<\tau^*\ll 1$ represents the smoothing parameter for approximation of the Dirac $\delta$ function (see Remark 1). 

\begin{table}
\caption{Model and numerical parameters used in numerical experiments.}
\label{table:1}

\begin{center}
\small
\begin{tabular}{ l || l }
 \hline
Model parameters & Numerical parameters
 \\
 \hline
$C_0= 0.02$         & $m$=200, 800  \\
$\kappa=0.3, \xi_{-} =0.05 \xi_{+}= 0.1$  &  $n$=250, 500\\
$\Delta t  =1/261$  &   $h$=0.01\\
$\sigma= 0.3$       &  $ \tau^{*}=0.005$\\
$r=0.011, q=0.008$  &   $k=T/m$\\
$T=1, E=50$         & $L=2.5$\\
 \hline
\end{tabular}
\end{center}

\end{table}

For the numerical parameters from Table~\ref{table:1}, we computed option values $V_{vtc}$ for several  underlying asset prices $S$. The prices were calculated by means of numerical solutions for both Bid and Ask option prices are shown in Table~\ref{table:3}. The Bid price $V_{Bid_{vtc}}$ is compared to the price $V_{BinMin}$ computed by means of the binomial tree method (cf. \cite{Kw}) with the constant lower volatility $\hat{\sigma}_{min}^2=\sigma^2(1-C_0\sqrt{\frac{2}{\pi}}\frac{1}{\sigma\sqrt{\Delta t}})$, whereas the upper bound price 
$V_{BinMax}$ corresponds to the solution with the higher constant volatility $\hat{\sigma}_{max}^2=\sigma^2(1-\underline{C_0}\sqrt{\frac{2}{\pi}}\frac{1}{\sigma\sqrt{\Delta t}})$. 

Similarly, as well as for the Ask price $V_{Ask_{vtc}}$ the lower bound $V_{BinMin}$ corresponds to the solution of the binomial tree method with the lower volatility $\hat{\sigma}_{min}^2=\sigma^2(1+\underline{C_0}\sqrt{\frac{2}{\pi}}\frac{1}{\sigma\sqrt{\Delta t}})$, whereas the upper bound $V_{BinMax}$ corresponds to the solution with the higher constant volatility $\hat{\sigma}_{max}^2=\sigma^2(1+{C_0}\sqrt{\frac{2}{\pi}}\frac{1}{\sigma\sqrt{\Delta t}})$.    

With regard to \cite{SeZ}, for a European style  option, one can derive the following lower and upper bounds by using the parabolic comparison principle:
\begin{equation*}
V_{\sigma_{min}}(S,t)\leq V_{vtc}(t,S)\leq V_{\sigma_{max}}(t,S),  \quad  S>0, t\in[0,T].
\end{equation*}    
In the case of American style options, analogous inequalities for the  numerical solution can be observed in Table~\ref{table:3}.

\begin{table}
\caption{Bid (top table) and Ask (bottom table) American call option prices  $V_{Bid_{vtc}}$ and  $V_{Ask_{vtc}}$ obtained from the numerical solution of the nonlinear model with variable transaction costs for different meshes. Comparison to the option prices $V_{BinMin}$ and $V_{BinMax}$ computed by means of the binomial tree method for constant volatilities $\sigma_{min}$ and $\sigma_{max}$.
}
\label{table:3}
\smallskip

\centering
\small
\begin{tabular}{r | r r r||r | r r r} 
\hline
& \multicolumn{3}{c||}{$n=250, m=200$} & & \multicolumn{3}{c}{$n=500, m=800$} \\
 $S$ & $V_{BinMin}$ & $V_{Bid_{vtc}}$ & $V_{BinMax}$ & $S$ & $V_{BinMin}$ & $V_{Bid_{vtc}}$ & $V_{BinMax}$ \\ [0.25ex] 
 \hline
 40 & 0.0320& 0.0513 & 1.3405 &40 &1.4511 & 1.6594 & 2.8670 \\  
 42 & 0.1075  & 0.3252 & 1.8846 &42& 2.0137  & 2.3869 & 3.6039 \\ 
 44 &0.2901 & 0.8232 & 2.5527 & 44&2.6979& 3.2309 & 4.4371 \\
 46 & 0.6535 & 1.5097 & 3.3483 & 46&3.5064 & 4.1868 & 5.3645 \\
48 & 1.2675 & 2.3859 & 4.2711 & 48&4.4382 & 5.2488 & 6.3833 \\
50 & 2.1740 & 3.4244 & 5.3175 & 50&5.4897 & 6.4133 & 7.4889 \\
52 & 3.3738 & 4.6126 & 6.4817 & 52&6.6553 & 7.6764 & 8.6772 \\
54 & 4.8304 & 5.9521 & 7.7555 &54 &7.9270 & 9.0342 & 9.9423 \\
56 & 6.4862 & 7.4377 & 9.1295 & 56&9.2959 & 10.4824 & 11.2798 \\
58 & 8.2809 & 9.0643 & 10.5943 & 58&10.7532 & 12.0179 & 12.6832 \\
 60 & 10.1635 & 10.8273 &  12.1397 & 60&12.2892 & 13.6385 & 14.1481\\ 
 \hline
\end{tabular}

\bigskip

\begin{tabular}{r | r r r||r | r r r} 
\hline
& \multicolumn{3}{c||}{$n=250, m=200$} & & \multicolumn{3}{c}{$n=500, m=800$} \\
 $S$ & $V_{BinMin}$ & $V_{Ask_{vtc}}$ & $V_{BinMax}$  & $S$ & $V_{BinMin}$ & $V_{Ask_{vtc}}$ & $V_{BinMax}$ \\ [0.25ex] 
 \hline 
 40 & 1.4511 & 1.6594 & 2.8670  &40 & 1.4420 & 1.6692 &2.8519 \\ 
 42 & 2.0137  & 2.3869 & 3.6039& 42 & 2.0027 & 2.3945 & 3.5870\\
 44 & 2.6979& 3.2309 & 4.4371 & 44 & 2.6851 & 3.2412 & 4.4187 \\
 46 & 3.5064 & 4.1868 & 5.3645&  46 & 3.4922 & 4.2134 &5.3450 \\
48 & 4.4382 & 5.2488 & 6.3833& 48 & 4.4231 & 5.2601 & 6.3627\\
50 & 5.4897 & 6.4133 & 7.4889 &50 & 5.4742 & 6.4300 & 7.4678 \\
52 & 6.6553 & 7.6764 & 8.6772 &52 & 6.6395 & 7.6922 & 8.6557 \\
54 & 7.9270 & 9.0342 & 9.9423 &54 & 7.9115 & 9.2167 & 9.9211\\
56 & 9.2959 & 10.4824 & 11.2798 &56 & 9.2812 & 11.0264 & 11.2586 \\
58 & 10.7532 & 12.0179 & 12.6832 &58 & 10.7393 & 12.2017 & 12.6628\\
 60 & 12.2892 & 13.6385 & 14.1481 & 60 & 12.2763 & 13.6505 &14.1283 \\ 
 \hline
\end{tabular}
\end{table}

In Table~\ref{table:4}, we present a comparison of results obtained by our method based on a solution to the Gamma variational inequality in which we considered constant volatilities $\sigma_{min}$ and $\sigma_{max}$ and those obtained by the well-known method based on binomial trees for American style call options (cf. \cite{Kw}). The difference in the prices is in the order of the mesh size $h=L/n$.              
\begin{figure}
\begin{center}
\includegraphics[width=0.45\textwidth]{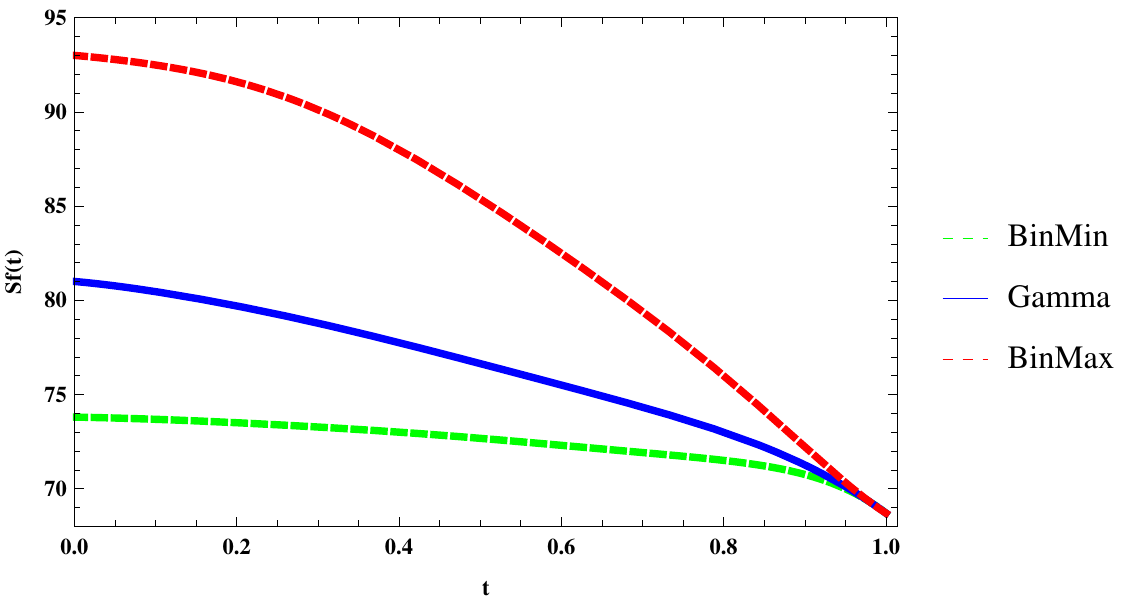}
\end{center}
\caption{The early exercise boundary function $S_{f}(t), t\in[0,T],$ computed for the model with variable transaction costs (dashed line Gamma) and comparison with early exercise boundary computed by means of binomial trees with constant volatilities $\sigma_{min}$ (bottom curve) and $\sigma_{max}$ (top curve).}
\label{Asset values}
\end{figure}

\begin{figure}
\begin{center}
\includegraphics[width=0.48\textwidth]{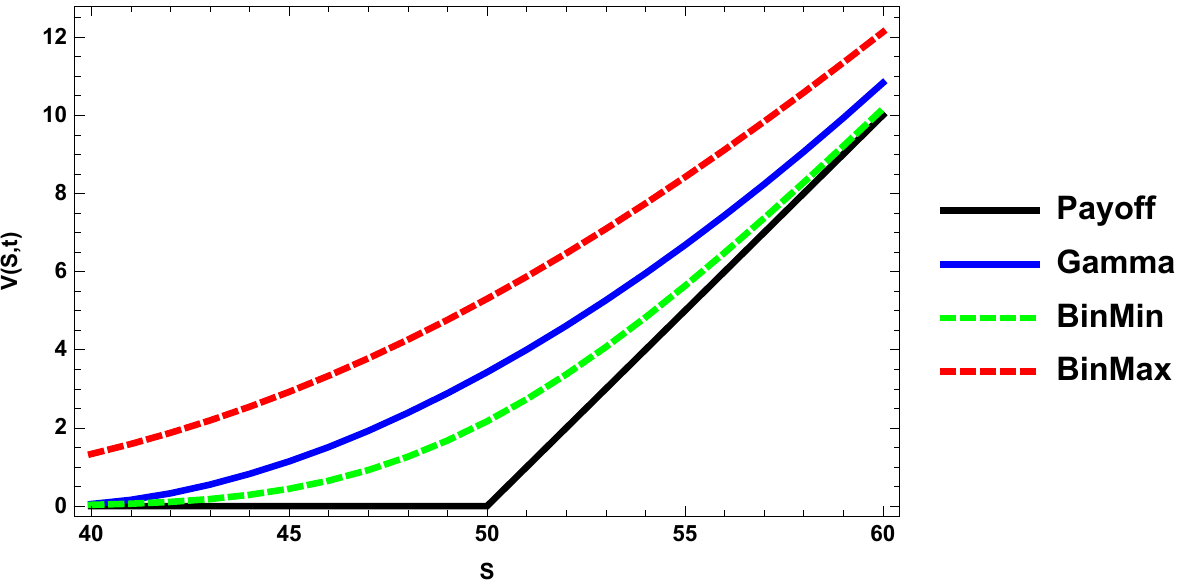}
\ \ 
\includegraphics[width=0.48\textwidth]{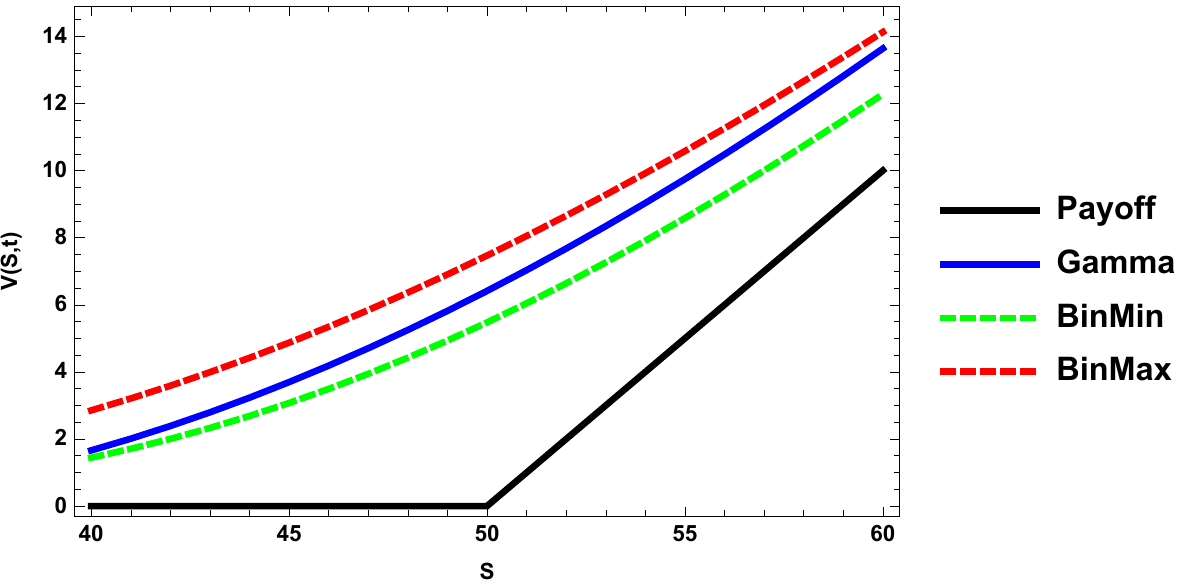}
\end{center}

\caption{A graph of American Bid (left) and Ask (right) call option prices $V(t,S), S\in[40,60],$ at $t=0$  computed by means of the nonlinear Black-Scholes model with variable transaction costs with the mesh size $n=500, m=800$ in comparison to solutions $V_{\sigma_{min}}, V_{\sigma_{max}}$ calculated by the binomial trees with constant volatilities $\sigma_{min}$ and $\sigma_{max}$.}
\label{option values}
\end{figure}

In Fig.~\ref{Asset values} we present the free boundary function $S_{f}(t)$ obtained by our method with a variable transaction costs function $C$ for bid option value compared to the binomial trees with $\sigma_{min}, \sigma_{max}$. In Fig.~\ref{option values} we plot the graphs of the solutions $V_{vtc}(t,S)$ at $t=0$ for both bid and ask prices. We also plot the prices obtained by the binomial tree method with the constant lower volatility $\sigma_{min}$ and the higher volatility $\sigma_{max}$, respectively. 

\begin{table}

\caption{Ask call option values $V_{Ask_{vtc}}$ of the numerical solution of the model under constant volatilities $\sigma=\sigma_{min}$ (left) and  $\sigma=\sigma_{max}$ (right) and  comparison to the prices computed by the Binomial tree method (with $n=100$ and $n=200$ nodes, respectively. 
}
\label{table:4}

\centering
\small
\begin{tabular}{r|rr|rr||r|rr|rr} 
 \hline
 & &\multicolumn{2}{c}{$\sigma=\sigma_{min}$} & &  && \multicolumn{2}{c}{$\sigma=\sigma_{max}$} &\\
 \hline
 & \multicolumn{2}{c}{$n=250, m=200$} & \multicolumn{2}{c||}{$n=500, m=800$} 
 & 
 & \multicolumn{2}{c}{$n=250, m=200$} & \multicolumn{2}{c}{$n=500, m=800$} \\ 
 $S$ & $V_{Ask_{vtc}}$ & $V_{BinMin}$ & $V_{Ask_{vtc}}$ & $V_{BinMin}$ & $S$ & $V_{Ask_{vtc}}$ & $V_{BinMax}$ & $V_{Ask_{vtc}}$ & $V_{BinMax}$ \\ [0.25ex] 
 \hline
 40  &  1.4737&1.4511 & 1.4634 & 1.4420 & 40& 2.8827 &2.8670&2.8663&   2.8519\\ 
 42    & 2.2417 &  2.0137 &  2.110&  2.002 & 42&3.6273 & 3.6039 &  3.5923 &   3.5870\\
 44  & 2.7156 & 2.6979  &    2.7025& 2.6851&44 & 4.4618& 4.4371 & 4.4067&  4.4187\\
 46   & 3.5287 &  3.5064 &  3.5193&  3.4922  &46& 5.3945 &  5.3645&  5.3561&   5.3450\\
48  &  4.4572&  4.4382 &  4.4498&  4.4231 &48& 6.4095 &  6.3833 &  6.3515&   6.3627\\
50  &  5.5019& 5.4897 &  5.4996& 5.4742 &50& 7.5002& 7.4889 &  7.4710&  7.4678\\
52  &  6.6993&  6.6553  &6.6684&  6.6395 &52&   8.7049&  8.6772 &  8.6682&   8.6557\\
54  &  7.9537&  7.9270 &7.9350&  7.9115 &54& 9.9765&  9.9423 &  9.9326&   9.9211\\
56  &  9.3367&  9.2959 & 9.3145&  9.2812 &56& 11.3071&  11.2798 &  11.2742&  11.2586\\
58  &  10.8015&  10.7532& 10.7683&  10.7393 &58& 12.7103&  12.6832 & 12.6790 &  12.6628\\
 60  &  12.3369&  12.2892& 12.3189&  12.2763 &60&   14.1640&  14.1481 &  14.1374&  14.1283\\ 
 \hline
\end{tabular}

\end{table}

\section{Conclusions}

\bigskip

In this paper we investigated a nonlinear generalization of the Black-Scholes equation for pricing American style call options assuming variable transaction costs for  trading the underlying assets. In this way, we presented  a model that addresses a more realistic financial framework than the classical Black-Scholes model. From the mathematical point of view, we analyzed a problem that consists of a fully nonlinear parabolic equation in which the nonlinear diffusion coefficient depends on the second derivative of the option price. Furthermore, for the American call option we transformed the nonlinear complementarity problem into the so-called Gamma variational inequality. We solved the Gamma variational inequality by means of the PSOR method and presented an effective numerical scheme for discretization the Gamma variational inequality. Then, we performed numerical computations for the model with variable transaction costs and compared the results with lower and upper bounds computed by means of the binomial tree method for constant volatilities. Finally, we presented a comparison between the respective early exercise boundary functions.

%\subsection*{Acknowledgements} This research was supported by the European Union in the FP7-PEOPLE-2012-ITN project STRIKE - Novel Methods in Computational Finance (304617), the project CEMAPRE MULTI/00491 financed by FCT/MEC through national funds and the Slovak research Agency Project VEGA 1/0780/15.        

\end{document}